\documentclass[10 pt]{article}
\usepackage{amsmath}
\usepackage{opex2}
\usepackage{graphicx}
\begin{document} 
\title{Characteristics of bound modes in coupled dielectric waveguides containing\\
negative index media} 
\author{Klaus Halterman, J. Merle Elson, and P. L. Overfelt}
\address{
Sensor and Signal Sciences Division, Naval Air Warfare Center, China Lake, CA 93555} 

\begin{abstract} 
We investigate the characteristics of guided wave modes in
planar coupled waveguides. In particular, we calculate the
dispersion relations for TM modes in which one or both
of the guiding layers consists of negative index media (NIM)-where the 
permittivity and permeability are both negative. We find
that the Poynting vector within the NIM waveguide axis can change sign and magnitude,
a feature that is reflected in the dispersion curves.
\end{abstract} 

\section{Introduction} 
With the continual refinement of laser arrays, and optical fibers,
the role of coupled waveguide systems in integrated optics has magnified in the
past few years.
The fabrication of arrays which use parallel dielectrics as guiding structures
has been demonstrated in millimeter-wave dielectric circuits. Furthermore,
the coupling of waves in planar dielectric sheet waveguides  provides a means
of reflectionless  signal transfer  from one waveguide to another.
For these systems, generally the constituent dielectric materials
are presumed to have positive values of the electric permittivity, $\epsilon$ and magnetic permeability, $\mu$.
Recently, however there has been interest
in researching materials which have the unusual property of both $\epsilon$ and $\mu$ being negative,
a concept first proposed  by Veselago \cite{veselago} long ago. 
These materials support electromagnetic wave
propagation in which the phase velocity is antiparallel to the energy flow, or group velocity. 
Furthermore, if both $\epsilon$, and $\mu$ are negative in a medium, the refractive index
has the extraordinary property of also being
negative, therefore such media are given the  name negative index media (NIM),
compared with the more familiar positive index media (PIM)-where both $\mu$ and $\epsilon$
are positive.
As discussed by Veselago \cite{veselago}, a multitude of other anomalous characteristics  of NIM follow from Maxwell's equations,
including reversal of the Doppler shift,  counter-directed Cherenkov radiation cone, and the refocusing of EM waves from a point source.
It was shown \cite{pendry} that a NIM slab can focus focus both near and far fields, attaining perfect resolution due to the 
amplification of  the evanescent waves. Subsequently however, the original ``perfect lens" model was modified \cite{ramak}  to overcome limitations
of imperfect materials by introducing a multilayer stack.
While the study of various NIM structures has intensified during the past
few years,
the number of theoretical works involving waveguides
with NIM components is less common.
The $S$ parameters and electromagnetic properties were investigated \cite{caloz}
numerically
for a $T$ junction NIM waveguide configuration, and a simple
plane two-layered waveguide was shown \cite{nefedov} to have a slow-wave factor that
tends to infinity at small frequencies.

On the experimental side, the fabrication of devices
constructed with
NIM is less prolific, as
there are no naturally occurring NIM in nature, and thus
one must artificially engineer a composite system having the
desired electromagnetic properties.
One such  medium was constructed \cite{smith} in 
which the effective permeability and effective permittivity were
shown to be simultaneously less than zero over a finite frequency band.
These structures that simulated a NIM were comprised of a periodic array of unit cells,
where each cell contained  two split metallic rings located on a dielectric panel
along with one copper wire.
The overall medium 
is considered homogeneous 
since the elements of each cell varied over length scales much smaller than
the wavelengths used in the experiments.
The experimental situation involving these unconventional materials is thus in
the early stages of development, 
and there exists a broad range of potential applications, including
optical and microwave devices.

It is the purpose of this paper to investigate the electromagnetic properties of a 
coupled planar  dielectric 
waveguide system. In order to study the
interaction between NIM and PIM configurations,
our method will allow for a
wide range of material parameters.
The paper is organized as follows. In Sec.~{\ref{method}}, we 
introduce the geometry and method used to 
calculate the relevant quantities of interest.
In Sec.~{\ref{results}}, we calculate the 
dispersion relations,
electromagnetic fields,
and energy flow characteristics
for differing
waveguide structures.
Finally in Sec.~{\ref{conclusion}}, we
summarize our results.

\begin{figure}
\centering
\includegraphics[width=3.5in]{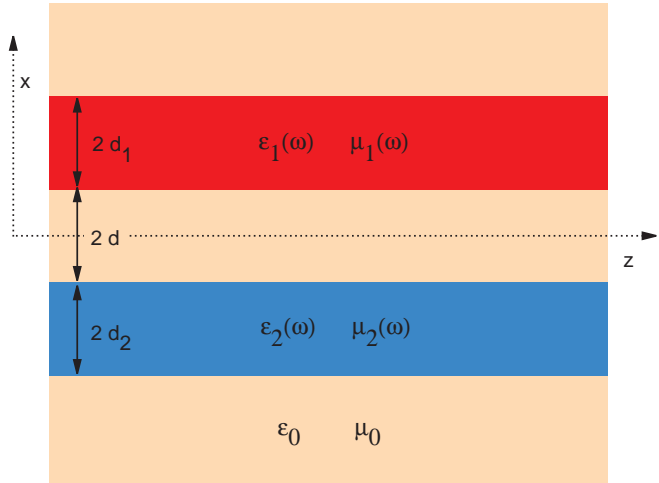}
 \caption{Coupled waveguide structure considered in this paper. The red and blue regions
correspond to the waveguide channels with dispersive material parameters. 
The permittivity and permeability in the channels can take positive or negative values. 
The free space parameters take the values
$\epsilon_0=\mu_0=1$.\label{pic}}
 \end{figure}

\section{Method}
\label{method}
\noindent
In this section we give a brief overview of  the model used for our coupled
waveguide system. The schematic of the planar dielectric waveguide is illustrated in Fig.~\ref{pic}.
The structure consists of two waveguide channels embedded in a medium 
characterized by the free space parameters $\epsilon_0$, and $\mu_0$.
The upper waveguide layer
of width $2 d_1$ has permittivity $\epsilon_1$ and permeability $\mu_1$,
and the lower guide of width $2 d_2$ has permittivity $\epsilon_2$ and permeability $\mu_2$.
The two channels are separated by a central film that  is $2 d$ in width and is divided in
the plane at $x=0$. The $z$ axis of the coordinate system coincides with the axis of the guide.

We are interested in guided wave modes, which are
electromagnetic waves that propagate
along the waveguide with a given phase and group velocity, intensity distribution, and
polarization.
These modes are relevant since the corresponding fields satisfy the wave equation throughout
the structure, and the appropriate boundary conditions. Furthermore,
each mode is characterized by its frequency $\omega$ and its wave vector $k_z$.
To proceed, we assume
a sinusoidal time dependence $\exp(-i \omega t)$ for the fields.
The dielectric media is linear and isotropic in the $y$ and $z$ directions.
The translational invariance in the $y$ and $z$ direction
allows, (for the case of
TM modes) one to write the magnetic field ${\bf H}$ as,
${\bf H}=\hat{y} h(x) \exp[i (k_z z - \omega t)]$,
where we have factored out the spatial variation of the field in 
the $x$ direction, and $k_z$ is the $z$ component of the wave vector.
Employing Maxwell's equations,
one can immediately write down the reduced wave equation 
that must be solved in each of the layers:
\begin{equation}
\label{wave}
\Bigl[\frac{\partial^2}{\partial x^2} -k_z^2+ \mu_i(\omega)\epsilon_i(\omega) \frac{\omega^2}{c^2} \Bigr]h(x) =0,
\qquad i=0,1,2,
\end{equation}
with an analogous expression for the ${\bf E}$ field.
In what follows, we 
consider material parameters in the 
NIM regions to be dispersive,
otherwise the
energy density would be negative \cite{veselago}. We take
the permittivity and permeability
to have the form,
\begin{equation}
\label{dis}
\epsilon_i(\omega)=1-\frac{\omega_e^2}{\omega^2},\qquad
\mu_i(\omega)=1-\frac{\omega_m^2}{\omega^2},
\end{equation}
where $\omega_e$, and  $\omega_m$ 
are the effective  electrical and magnetic plasma frequencies \cite{pendry}, respectively.
To simplify the model, we have neglected damping.
Elementary solutions to Eq. (\ref{wave}) admit the following wave vectors,
\begin{equation}
k_i = \sqrt{\frac{\epsilon_i (\omega) \mu_i (\omega) \omega^2}{c^2}-k_z^2} \qquad i=0,1,2,
\end{equation}
where the wave vectors are labeled according to the respective values of 
the permeability  and permittivity in each region depicted in Fig.~\ref{pic}.
It is clear that the longitudinal wave vector $k_z$, common to all the layers, serves
as a connection to constrain the transverse components.

\section{Results \label{results}}
\subsection{NIM Waveguides}
\begin{figure}[t!]
\centering
\includegraphics[width=4in]{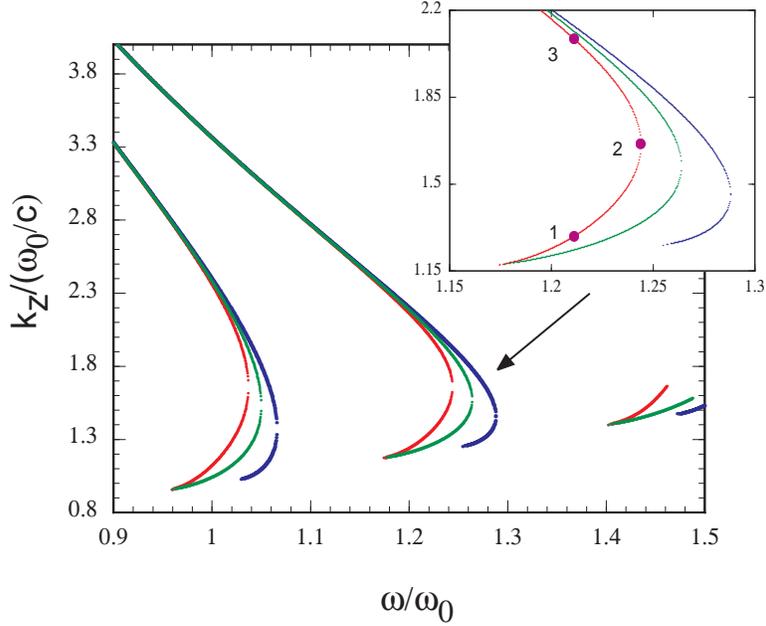}
 \caption{Dispersion curves for waveguide structure with material
parameters given by Eqs.~(\ref{dis}). 
The widths of the
middle layer and the two waveguide channels are given by $d=0.1 \lambda_0$, and
$d_1=0.2 \lambda_0$ respectively, where 
$\omega_0 = 2 \pi c/\lambda_0$. 
The red curve corresponds to the symmetric modes,
and the blue curve is for the asymmetric modes.
The central green curves correspond to the dispersion relation for a single NIM waveguide
of thickness $2 d_1$.
The inset is a magnification of the same quantity localized around one
of the bending features. The numbers label points of interest as discussed in the text.
\label{nim}}
 \end{figure}
We consider first the case where 
the permittivity and permeability are both simultaneously negative in
each waveguide core, and
each of the  guiding layers
has the same width $d_1=d_2$ (see Fig.~\ref{pic}).
To derive the TM modes, we proceed to seek elementary solutions of Eq. (\ref{wave})
that do not radiate away from the system. Upon invoking the continuity of $E_z$ and $H_y$
at the interfaces, we arrive at the following eigenvalue equation or
dispersion relation for the symmetric and antisymmetric modes:
\begin{equation}
\label{disp2}
\pm e^{-2 k_0 d} \sin(2 k_1 d_1)-\sin(2 k_1 d_1-2 \psi)=0,
\end{equation}
where we have introduced a phase angle $\psi$,
\begin{equation}
\psi \equiv \arctan\Bigl(\frac{k_0 \epsilon_1}{k_1 \epsilon_0}\Bigr).
\end{equation}
We show in Fig.~\ref{nim}, 
the dispersion relation for the  case where all the  dielectric layers are
made of NIM. We plot $k_z$ normalized by a design frequency $\omega_0/c$  versus the dimensionless
frequency $\omega/\omega_0$. In the following, we take $\omega_e=2.34 \omega_0$,
and $\omega_m=1.98 \omega_0$.
All curves are bound by the light line, $k_z=(\omega/c)\sqrt{\epsilon_0 \mu_0}$,
and the limiting line $k_z=(\omega/c)\sqrt{\epsilon_1(\omega) \mu_1(\omega)}$. Thus, the ``phase-space" available
for guided wave modes is contained within the area spanned by these two lines, and increases rapidly
upon decreasing $\omega$, due to material dispersion.
The figure illustrates the dispersion diagram with a finite number of bands, which arise
from the geometrical constraint placed upon the fields from the bounding surfaces.
Since the fields outside the waveguide channels decay over the length scale $k_0^{-1}$,
regions of the dispersion curves near the light line correspond to a larger decay length,
thus the fields reside mainly outside the guiding layer. The opposite is true for the $\omega$
and $k_z$ dispersion pair which reside in the vicinity of the line $k_z=(\omega/c)\sqrt{\epsilon_1(\omega) \mu_1(\omega)}$.
The effect of increasing the coupling (decreasing $d$) 
between the two waveguide cores
has the  trend 
of increasing the divide between the two modal bands with different symmetry. 
Although not visible in the plot, the zeroth order TM mode is cutoff in frequency,
in contrast to the usual PIM case where there is no frequency cutoff.
By way of comparison, we also show in Fig.~\ref{nim}, the dispersion curve for a single NIM waveguide.
It is apparent that the  symmetric and antisymmetric modes are centered about the 
single waveguide case.
One of the most dramatic features of the dispersion relation is the 
bending back of  the  bands close to the light line. This is a behavior 
seen in general whenever a waveguide channel consisting of NIM is in 
proximity to a material with PIM characteristics.
The inset of Fig.~\ref{nim} shows a close up of the second modal branch.
Some salient features to be noted are the change in slope of the dispersion curve,
varying from  positive to negative as one cycles through the points $1-3$ respectively.
We thus find some 
peculiarities in the dispersion diagram that 
clearly deviate from the well understood
coupled waveguide system with conventional
dielectric material. As will be
discussed below, there is an abrupt reversal of  energy flow
in the $z$-direction 
between the two different media, and this quite naturally is reflected in the allowed
guided wave modes.
In order to further investigate this anomalous  behavior of the energy flow in the
vicinity of a bend,
it is necessary to calculate the Poynting vector, which requires the 
full electromagnetic 
fields.

\begin{figure}[t!]
\centering
\includegraphics[width=3.5in]{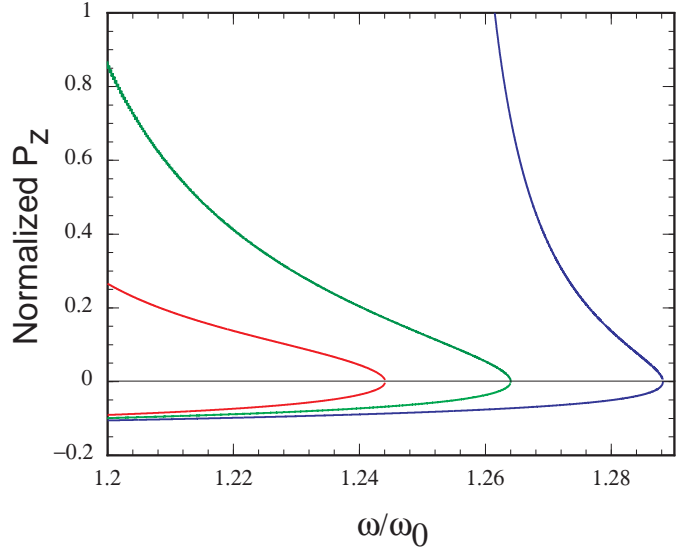}
 \caption{The normalized $z$ component of power,
$P_z$, is plotted as a function of dimensionless frequency $\omega/\omega_0$.
The color labeling of the  curves coincides with the dispersion
curves displayed in the inset of Fig.~\ref{nim}. For
comparison purposes, the green curve given by Eq.~(\ref{pz}),
is multiplied by a factor of 2.
\label{poyntavg}}
 \end{figure}

Having obtained the set of $k_z$ and $\omega$ that  are solutions to  Eq. (\ref{disp2}),
the spatially varying electromagnetic fields can be straightforwardly calculated from Eq.(\ref{wave}).
To demonstrate this, we write down the magnetic fields 
consistent with the dispersion relation in Eq.(\ref{disp2}),
\begin{subequations}\label{bogo}
\begin{align}  
h(x)&=\frac{\cos(\psi)}{ \cos(k_1 d_1-\psi)} e^{-k_0 (x-L)}, \qquad x\ge L,\\
&=\frac{\cos[k_1(x-L)+\psi]}{\cos(k_1 d_1-\psi)}, \qquad d \le x \le L,\\
&=\frac{ \cos(2 k_1 d_1-\psi)}{\cos(k_1 d_1-\psi) \cosh(k_0 d)}\cosh(k_0 x),\qquad |x| \le d,\\
&=\frac{\cos[k_1(x+L)-\psi]}{\cos(k_1 d_1-\psi)}, \qquad -d \le x \le -L,\\
&=\frac{\cos(\psi)}{ \cos(k_1 d_1-\psi)}  e^{k_0 (x+L)}, \qquad x\le -L,
\end{align} 
\end{subequations}
where $L \equiv 2 d_1 + d$, and for
brevity we show only the symmetric field, 
noting that the antisymmetric fields are calculated similarly.
\begin{figure}
\centering
\includegraphics[width=3.6in]{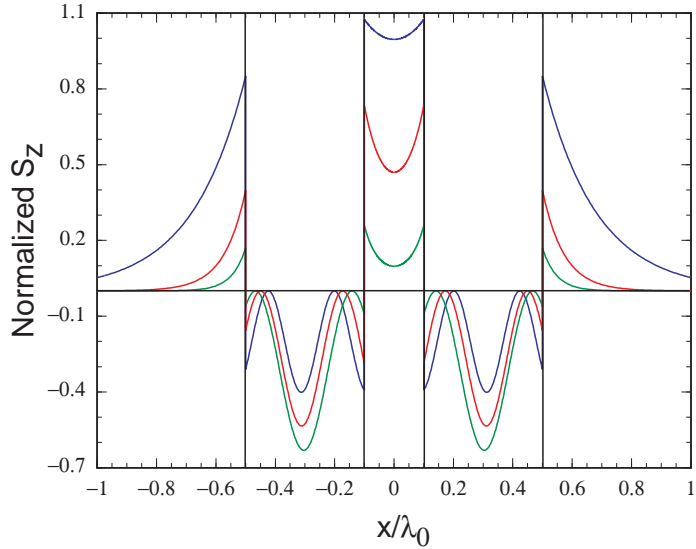}
 \caption{The normalized $z$-component of the Poynting vector $S_z$
as a function of dimensionless position. The continual decline in 
amplitude of $S_z$ in the central film and
the outer two layers from blue to green coincides (in order)
with points 1-3 in the inset of Fig.~\ref{nim} respectively.
Meanwhile, the two waveguides
have relatively little shift in amplitude. 
The vertical lines serve as guides to the eye.
\label{poynt}}
 \end{figure}
The magnetic field distribution in Eq. (\ref{bogo}) reflects the fact that the fields are oscillatory within
the two waveguide cores, and exponentially decaying in the outer regions.
For a given set of material parameters, Eqs. (\ref{bogo}) supplemented with Eq. (\ref{disp2}) completely
determine (apart from an unimportant overall constant amplitude factor) the electromagnetic properties of the structure.
We are ultimately interested in the
time-averaged  power flow density however, as 
given by the real part of the Poynting vector: ${\bf S}=[c/8 \pi]({\bf E \times H^*})$. 
Upon calculating the electric field distribution from Eq. (\ref{bogo}) by taking the appropriate
derivatives via  Maxwell's equations, we find that
the transverse component,  $S_x$ is imaginary, and thus does not contribute to the time-average flux of energy.
The $z$ component however does contribute, and is given by,
\begin{equation}
\label{sz}
S_z(x;k_z,\omega) = \frac{c^2}{8 \pi} \frac{k_z}{\omega \epsilon_i(\omega)} [h(x)]^2,\qquad i=0,1,2.
\end{equation}
In this instance, the direction of ${S_z}$ depends upon the sign of the permittivity of the medium,
and is discontinuous across the boundaries. On the other hand,
the longitudinal wave vector $k_z$ remains continuous across each layer,
as required by translational invariance. 
To illustrate the spatial dependence of $S_z$, we show in Fig.~\ref{poynt},
the time averaged Poynting vector throughout the waveguide system. Each of the three
curves corresponds to an allowed $k_z$ and $\omega$ pair identified
by the three points within the inset of Fig.~\ref{nim}. The most visible behavior
is the rapid decline in $S_z$ in the regions characterized by material parameters $\epsilon_0$
and $\mu_0$, while the two NIM waveguide channels show relatively little variance. 
The blue curve reveals that more total energy is flowing in the 
central and two outer layers, while  the
parameters corresponding to the green curve shows that
the net negative $S_z$ in the waveguide regions is
greater overall. 
This trend continues at other points, so that
the sign of the slope of the dispersion relation 
correlates with that of  the net energy flow.
Since  $d \omega/d k_z$ changes sign, there must  be points
in the dispersion curve that result in
$S_z$  in the middle and outer layers exactly countering
$S_z$ in the two waveguide channels.
This is shown by the intermediate red curve in Fig.~\ref{poyntavg}.
For this special point on
the dispersion curve, $d \omega/d k_z=0$ (see point 2, inset of Fig.~\ref{nim}), and the net $S_z$ is zero.

The interesting properties
of the net energy flow in our coupled waveguide system can be investigated further 
by calculating the time average of the power flow $P_z$ across all the layers. In
order to do this, we
spatially integrate Eq.(\ref{sz}) over all of the spatial region encompassing the system.
For the single layer case, $P_z$ can be simply expressed as
\begin{equation}
\label{pz}
P_z(k_z,\omega)\equiv \int_{-\infty}^{\infty} dx S_z(x;k_z,\omega)=
\frac{c^2}{8 \pi}\frac{k_z d_1}{\omega}\Bigl\lbrace\frac{\cos^2(k_1 d_1)}{\epsilon_0 k_0 d_1}+
\frac{1}{\epsilon_1}\bigl[1+\frac{\sin(2 k_1 d_1)}{2 k_1 d_1}\bigr]\Bigr \rbrace.
\end{equation}
Figure \ref{poyntavg} shows the power across the whole system as a function
of modal frequency. 
We see that the single layer result is always between
the symmetric and antisymmetric cases.
The positive $P_z$ curves correspond to
the lower branch in the dispersion diagram, below the point where
 $d \omega/d k_z = 0$ (see Fig.~\ref{nim}). Likewise, the  negative $P_z$ curves 
correlate with
those $\omega$, $k_z$ pairs
above that point.
Thus  a clear relationship exists between
the net power and the slope of the dispersion curves:
$P_z$  is always negative for values of $k_z$ and $\omega$ that
satisfy $d \omega/d k_z <0$, and is positive
when  $d \omega/d k_z >0$.
For those values of $k_z$ and $\omega$ where  $P_z=0$,
we again find consistency with the dispersion curves, where at those same points 
$d \omega/d k_z = 0$. This result follows from the exact 
cancellation of negatively directed energy flow in the 
waveguides with the positive contributions from the outer PIM layers.

\subsection{NIM/PIM  Waveguides}
\begin{figure}
\centering
\includegraphics[width=3.5in]{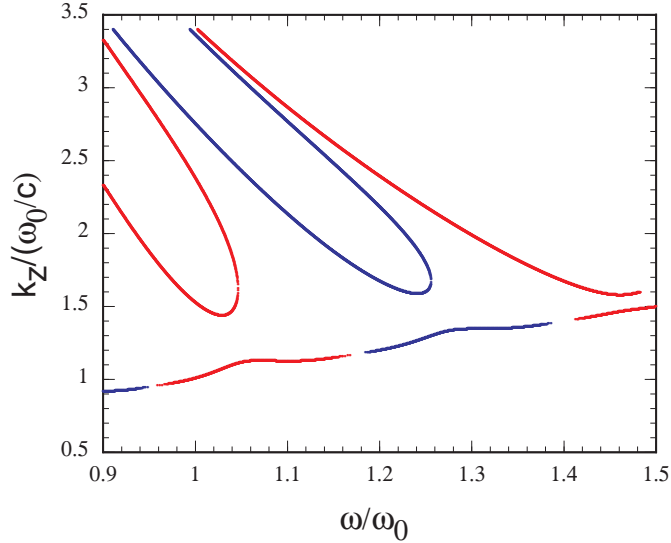}
\caption{Dispersion relation for the coupled waveguide structure 
consisting of a NIM channel and a PIM channel. The widths of the
middle layer and the two guide channels are given by $d=0.1 \lambda_0$, and
$d_1=0.2 \lambda_0$ respectively.  The red (blue) curves correspond to the plus (minus)
sign in Eq. (\ref{disp3}).\label{disp4}}
\end{figure}

Next, we consider the case where the two waveguide channels 
possess opposite material characteristics, i.e.,
$\epsilon_1=-\epsilon_2$, and $\mu_1=-\mu_2$.
After a lengthy calculation, the  dispersion relation can be expressed as:
\begin{equation}
\label{disp3}
\sqrt{1-e^{-4 k_0 d}}\sin(2 k_1 d_1) \pm \sin(2 \psi)=0.
\end{equation}
\noindent In Fig.~\ref{disp4} we  
present the relationship between $k_z$  and $\omega$ for the present asymmetric 
waveguide configuration. The two differently colored curves shown correlate to the choice of
sign in Eq. (\ref{disp3}). Since there is no longer any symmetry in the $x$ direction, the overall behavior 
of the dispersion curves is more complex than the previous case, where 
for a given frequency range, 
each waveguide had identical negative
permittivity and permeability values. 
The figure 
illustrates the unusual change in sign of $d \omega/d k_z$
that occurs when traversing the 
two visible U-shaped branches.
To proceed with the calculation of the Poynting vector, we again
calculate the electromagnetic fields within the five different layers.
We find,
\begin{subequations}\label{bogo2}
\begin{align}  
h(x)&=\frac{ \sin[2(k_1d_1+\psi)]}{\sin (2k_1 d_1)}e^{-k_0(x-2 d)} , \qquad x\ge L,\\
&=\frac{e^{-k_0 (L+2 d)}\sin(2 k_1 d_1)}{\cos (\psi) \sin[2(k_1 d_1 - \psi)]} \cos[k_1(x-L)+\psi] , \qquad d \le x \le L,\\
&=\frac{e^{-k_0 L}}{\sin (2 \psi)} \left\lbrace e^{k_0 (x+d)} \sin[2(k_1 d_1+\psi)]-
e^{-k_0 (x+d)} \sin[2(k_1 d_1)]\right\rbrace, \,\,\,  |x| \le d,\\
&=\frac{e^{-k_0 L}}{\cos (\psi)} \cos[k_1(x+L)+\psi] ,\qquad  -d \le x \le-L,\\
&={e^{k_0 x}}, \qquad x\le -L.
\end{align} 
\end{subequations}
\noindent From the plot in Fig.~\ref{disp4}, we select four points $(k_z, \omega)$ on 
the dispersion curves intersected by the $\omega/\omega_0 = 1.24$ line. 
Inserting these  points into Eqs. (\ref{bogo2}) and then into (\ref{sz}) results in the 
\begin{figure}[t!]
\centering
\includegraphics[width=3.6in]{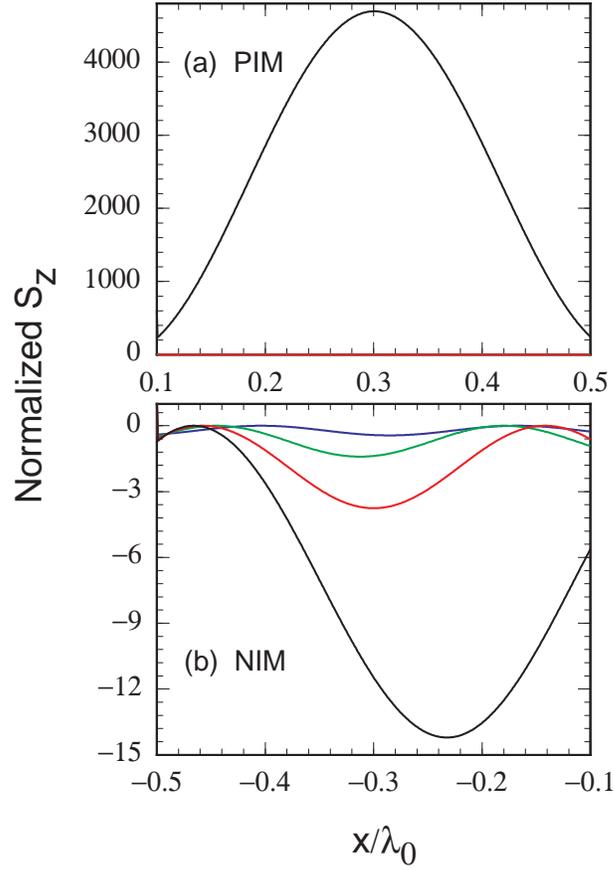}
 \caption{The spatial dependence of the (normalized)
Poynting vector  $S_z$, in both waveguide cores, as a function of $x/\lambda_0$.
The top panel (a) depicts $S_z$ in the PIM guide,  and the bottom
panel (b) corresponds to the NIM guiding layer.
Energy flow is always negative in the NIM waveguide, and positive in the PIM waveguide.
For all four curves, $\omega/\omega_0 = 1.24$ and $k_z/(\omega_0/c)  = 1.27$ (blue), 
1.59(green), 1.89 (red), and 2.22 (black).
Upon comparing both panels, it is evident that for   $k_z/(\omega_0/c)  = 2.22$, the PIM
waveguide layer bears most of the energy flux from
the electromagnetic fields.
\label{poynt2}}
 \end{figure}
normalized spatially dependent energy flow that is shown in Fig.~\ref{poynt2}. 
The upper panel in Fig.~\ref{poynt2}(a) depicts a relatively
flat red line, which is actually three curves merged together, along with a
dominant energy distribution (black curve) for the parameter values  $\omega/\omega_0 = 1.24$,
and $k_z/(\omega_0/c)  =2.22$. Thus a feature arises in which
three out of the four wave vector values have negligible contributions to $S_z$ in this region.
Proceeding to the lower panel (b), we exhibit $S_z$ for the same parameter values
as in (a), except we consider now the NIM region.
In addition to the energy flow reversing direction, all four $k_z$ and
$\omega$ pairs are now visible and have a
relatively complicated asymmetry when 
compared to the case when the waveguide structure is physically symmetric, 
as in Fig.~\ref{poynt}, where the guiding layers are both NIM media. 
It is clear therefore,
that there can be cases where the energy flow is dominant in either the NIM 
or PIM slab, or is shared between both slabs.

\section{Conclusion \label{conclusion}}
In this paper we have examined the geometrical and material dispersion properties of
a coupled planar waveguide structure that contains negative index media. We calculated the
TM guided wave modes and found a variety of unconventional results. For the case where both coupled waveguides have NIM characteristics and are embedded in PIM media, 
the dispersion relation revealed clear deviations 
from the well understood case involving positive material parameters. We found that due
to the reversal of energy flow along the propagation direction, the dispersion curve reflected
a striking behavior in curvature. This followed ultimately from the correlation between 
the net power in the system and the slope $d \omega/d k_z$. Furthermore, 
decreasing the relative distance between each wave channel has the effect
of further splitting the symmetric and antisymmetric dispersion curves.
Finally, we presented results for a more complicated asymmetric waveguide structure
comprised of both  NIM and PIM guiding layers. In this case, the Poynting vector and 
dispersion characteristics exhibited nontrivial behavior. In particular,
we found that there are cases when the energy flow reverses direction when going
from the NIM to PIM layer, and that the relative distribution of power
varied dramatically between the two, depending on the particular guided wave mode
under consideration.
\section*{Acknowledgments}
This work was supported in part by a grant of high performance computing time
           from the DoD NAVO Major Shared Resource Center.

\end{document}